\documentclass[aps,prb,twocolumn,floatfix,superscriptaddress]{revtex4}
\usepackage{amsmath}
\usepackage{amssymb}
\usepackage{graphicx}
\usepackage{dcolumn}
\usepackage{bm}

\begin{document}

\title{Optimal laser-control of double quantum dots}
\author{E. R{\"a}s{\"a}nen}
\email[Electronic address:\;]{esa@physik.fu-berlin.de}
\affiliation{Institut f{\"u}r Theoretische Physik,  
Freie Universit{\"a}t Berlin, Arnimallee 14, D-14195 Berlin, Germany}
\affiliation{European Theoretical Spectroscopy Facility (ETSF)}
\author{A. Castro}
\affiliation{Institut f{\"u}r Theoretische Physik,  
Freie Universit{\"a}t Berlin, Arnimallee 14, D-14195 Berlin, Germany}
\affiliation{European Theoretical Spectroscopy Facility (ETSF)}
\author{J. Werschnik}
\affiliation{JENOPTIK Laser, Optik, Systeme GmbH, G{\"o}schwitzer
  Str. 25, D-07745 Jena, Germany}
\affiliation{Institut f{\"u}r Theoretische Physik,  
Freie Universit{\"a}t Berlin, Arnimallee 14, D-14195 Berlin, Germany}
\affiliation{European Theoretical Spectroscopy Facility (ETSF)}
\author{A. Rubio}
\affiliation{Departamento de F{\'i}sica de Materiales, Facultad de Qu{\'i}micas
Universidad del Pa{\'i}s Vasco, Centro Mixto CSIC-UPV, Donostia International
Physics Center (DIPC), E-20018 Donostia-San Sebasti{\'a}n, Spain}
\affiliation{Institut f{\"u}r Theoretische Physik,  
Freie Universit{\"a}t Berlin, Arnimallee 14, D-14195 Berlin, Germany}
\affiliation{European Theoretical Spectroscopy Facility (ETSF)}
\author{E. K. U. Gross}
\affiliation{Institut f{\"u}r Theoretische Physik,  
Freie Universit{\"a}t Berlin, Arnimallee 14, D-14195 Berlin, Germany}
\affiliation{European Theoretical Spectroscopy Facility (ETSF)}

\date{\today}

\begin{abstract}
Coherent single-electron control in a realistic
semiconductor double quantum dot is studied theoretically. 
Using optimal-control theory we show that the energy 
spectrum of a two-dimensional double quantum 
dot has a fully controllable transition line.
We find that optimized picosecond laser pulses generate 
population transfer at significantly higher 
fidelities ($>0.99$) than conventional sinusoidal 
pulses. Finally we design a robust and fast 
charge switch driven by optimal pulses that are 
within reach of terahertz laser technology.
\end{abstract}

\maketitle

%%%%%%%%%%%%%%%%%
% INTRODUCTION %
%%%%%%%%%%%%%%%%
Double quantum dots (DQDs), i.e., coupled two-dimensional (2D) 
electron traps, have been under recent and extensive studies 
both experimentally~\cite{pettaPRL,pettascience,gorman} and 
theoretically.~\cite{forre,harju,wensauer} The main interest in 
DQDs arises from their potential for solid-state
quantum computation that could be achieved in principle
by rapidly switching voltages of electrostatic gates.
The gates permit to tune at will the system geometry and
hence the electronic properties of DQDs.
Coherent manipulation of a single charge~\cite{pettaPRL} 
and coupled spins~\cite{pettascience} has already been 
achieved, and recently a coherence time of $\sim 200$ 
ns was obtained for a well isolated silicon 
DQD.~\cite{gorman}
Theoretical studies on single-electron transport 
inside the DQD driven by linear switches and linearly 
polarized continuous waves (CWs) were reported very 
recently.~\cite{forre}
In the latter case the transport is rather sensitive to 
possible anharmonicity of the potential and limited
to uncoupled dots far apart from each other.
Electron control in DQDs has been studied also
using genetic algorithms~\cite{garcia} as well as 
rotating-wave and resonant approximations leading
to a reduction to a three-level system.~\cite{kosionis}
To the best of our knowledge, however, a {\em general}
$N$-level control scheme by using direct external electric 
fields has not been introduced for 2D-DQDs until now.

In this paper we discuss the controllability 
criteria for single-electron states of DQDs by
means of external laser pulses. We show that at certain 
interdot distances some of the single-electron states 
allow full 
population transfer from the ground state to those states.
We apply quantum optimal-control theory (OCT)~\cite{oct}
which yields the optimal laser pulses for
predefined transitions. We obtain high occupations ($\gtrsim 99\%$)
of the target states in a realistic DQD in a few picoseconds,
which is well in the coherent regime. If the initial 
and final states are chosen to have full localization of the 
electron in one or the other dot, this scheme 
enables rapid and controlled transport which is not
sensitive to the interdot distance or to the inevitable 
anharmonicities in the confining potential.

%%%%%%%%%%%%%%%%%%%
% MODEL & METHODS %
%%%%%%%%%%%%%%%%%%%
In the static 2D Hamiltonian,
${\hat H}_0=-\left(\partial_x^2+\partial_y^2\right)/2+V_{\rm c}(x,y)$,
the external potential describing the
DQD is, in its most common form,~\cite{wensauer} given by
\begin{equation}
V_{\rm c}(x,y)=\frac{\omega_0^2}{2}\,{\rm min}\left[(x-\frac{d}{2})^2+y^2,(x+\frac{d}{2})^2+y^2\right],
\label{pot}
\end{equation}
where $d$ is the distance between the potential minima,
and $\omega_0=0.5$ is the confinement strength with 
a typical value for DQDs.
We apply the effective-mass 
approximation for electrons moving in GaAs and set
the effective mass to $m^*=0.067m_e$ and the
dielectric constant to $\kappa=12.7\epsilon_0$.
The energies, lengths, and times are given in effective
atomic units (a.u.):
${\rm Ha}^*=(m^*/\kappa^2){\rm Ha}\approx 11.30$ meV, 
$a_0^*=(m^*/\kappa)a_0\approx 10.03$ nm, and 
$u^*_t=\hbar/{\rm Ha}^*\approx 58.23$ fs, respectively.

\begin{figure}
\includegraphics[width=0.9\columnwidth]{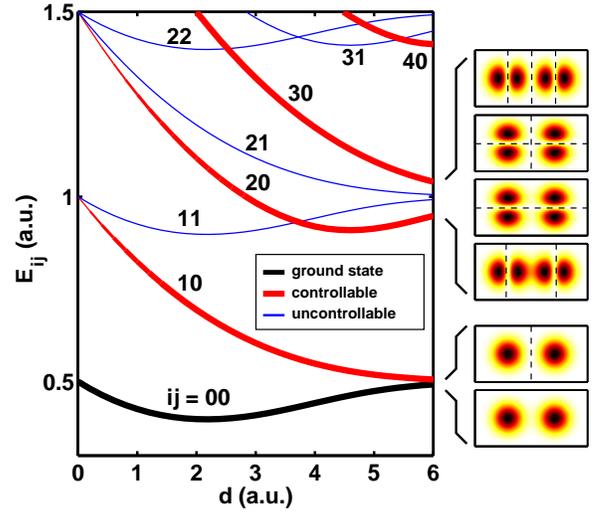}
\caption{(color online). Left panel: Lowest eigenenergies of a 
double quantum dot with $\omega_0=0.5$ as a function of the interdot distance.
Black, red (thick), and blue (thin) curves mark the ground
state, controllable states, and uncontrollable states, 
respectively.
Right panel: Densities of six lowest eigenstates at $d=5$.
The dashed lines mark the nodes of the wave functions.}
\label{spectrum}
\end{figure}

In Fig.~\ref{spectrum}
we show the lowest energy levels 
%from the static Schr{\"o}dinger equation 
%$\hat{H_0}\psi_{ij}=\epsilon_{ij}\psi_{ij}$ 
as a function of $d$ (left panel) as well as densities of 
six lowest eigenstates at $d=5$ (right panel). The $d=0$ limit corresponds
to the well-known shell structure of a single 2D harmonic 
oscillator (HO), where the energy levels are $n$-fold degenerate 
at energies $n\omega_0$ ($n=1,2,\ldots$). Increasing $d$ leads to 
lifting of the degeneracies and generates crossings and avoided crossings 
between the energy levels. In the weak-coupling limit 
$d\rightarrow \infty$ the bundling of the levels at $n\omega_0$ is 
restored. In this limit, the energies are $2n$-fold
degenerate corresponding to two (uncoupled) HOs. 
We label the states in the DQD 
as $|\psi_{ij}\big>=|ij\big>$, where $i=0,1,\ldots$ denotes the 
$(i+1)$th bundle of 
states at $d=0$, and $j=0,1,\ldots$ denotes 
the $(j+1)$th state in each bundle. As visualized in the 
right panel of Fig.~\ref{spectrum}, index $i$ also corresponds 
to the number of nodes in the wave function. We show below
that our labeling yields simple rules for controllable and
uncontrollable states as a function of $d$.

It is a well-known fact in control theory 
that an infinite-level single HO 
is not controllable in the dipole 
approximation.~\cite{qdcontrol} This is in contrast
with the truncated HO which
(in most cases) satisfies the 
controllability criteria.~\cite{schirmer}
In qualitative arguments, the uncontrollability
of a single HO ($d=0$) in the dipole approximation stems from
the equidistant single-electron level spacings. 
Using the above labeling of the states, the nonvanishing
dipole-matrix elements $\left<ij|\hat{\mu}|kl\right>$
between the HO states ($d=0$) are
\begin{eqnarray}
|\left<i,j|\hat{\mu}|i+1,j+1\right>| & = & \sqrt{2(j+1)}, 
\label{condition1}
\\
|\left<i,j|\hat{\mu}|i+1,j\right>| & = & \sqrt{2(i-j+1)},
\label{condition2}
\end{eqnarray}
where ${\hat {\mathbf \mu}}=-e\mathbf{r}$ is the dipole operator.
Eq.~(\ref{condition1}) holds for {\em all values of} $d$, and
the corresponding energy-level spacings remain constant.
Hence, transitions $ij\rightarrow (i\pm 1)(j\pm 1)$ remain
uncontrollable. On the other hand, the elements given in the 
LHS of Eq.~(\ref{condition2}) change as a function of $d$,
as well as the level spacings. The behavior already suggests 
that transitions $ij\rightarrow (i\pm 1)j$ become controllable 
when $d$ is increased from zero. Our calculations below confirm
this prediction. Starting with the ground state
$|00\big>$, the controllable transition line is then
$|00\big>\rightarrow |10\big> \rightarrow |20\big> \ldots$ (see the red curves
in Fig.~\ref{spectrum}). In the weak-coupling limit 
($d\rightarrow\infty$), however,
the LHS of Eq.~(\ref{condition2}) becomes
\begin{equation}
\lim_{d\rightarrow \infty}|\left<i,j|\hat{\mu}|i+1,j\right>| = \left\{ \begin{array}{ll}
\infty & \textrm{if $i-j$ is even}\\
\sqrt{2(i-j)} & \textrm{if $i-j$ is odd},
\end{array} \right.
\end{equation}
so that only the trivial transitions between the 
degenerate gerade and ungerade states are possible.
In the large-$d$ regime, however, breaking the interdot 
symmetry leads to the possibility of charge transport 
between the dots (see below).
%These transitions are not of relevance for the 
%quantum-information applications of DQDs.

Next we apply OCT in order to find optimal laser pulses
for transitions from the initial state $|\Phi_{\rm I}\big>=|\Psi(t=0)\big>$
to the target state $|\Phi_{\rm F}\big>=|\Psi(t=T)\big>$ in a fixed time interval $T$.
%We note that OCT approach is an efficient alternative to using
%simple CWs with frequencies $\omega_{\rm IF}=(E_{\rm F}-E_{\rm I})$
%that achieve accurate occupations only in the limit $T\rightarrow\infty$.
In the OCT formalism we maximize the overlap 
$|\left<\Psi(T)|\Phi_{\rm F}\right>|^2$ while minimizing the fluence of the
laser pulse. The control equations are~\cite{ring}
\begin{eqnarray}
i\partial_t\Psi(t) & = & \hat{H}\Psi(t), \,\,\Psi(0)=\Phi_{\rm I},\label{tdse} \\
i\partial_t\chi(t) & = & \hat{H}\chi(t),
\,\,\chi(T)=\Phi_{\rm F}\left<\Phi_{\rm F}|\Psi(T)\right>,\\
{\boldsymbol \epsilon}(t) & = & -\frac{A(t)}{\alpha}{\rm Im}\,\left<\chi(t)|\hat{\mu}|\Psi(t)\right>,\label{pulse}
\end{eqnarray}
where Eq.~(\ref{tdse}) is the time-dependent Schr{\"o}dinger equation
with $\hat{H}=\hat{H}_0-\hat{\mu}{\boldsymbol \epsilon}(t)$, and
$\chi(t)$ is the Lagrange multiplier. The optimal 
laser pulse ${\boldsymbol \epsilon}(t)$ is provided at the 
end of the iterative procedure.~\cite{zhu} We point out that
the initial pulse (zeroth iteration) is sinusoidal and has 
both {\em x} and {\em y} components, whereas the converged 
optimal pulse is always found to be polarized in the {\em x} 
direction, i.e., ${\boldsymbol \epsilon}(t)=\epsilon(t)\hat{x}$.
In Eq.~(\ref{pulse}) we 
choose a sinusoidal pulse envelope $A(t)=\sin^2(\pi t/T)$,
and restrict the pulse intensity by a penalty factor $\alpha$.
Unless stated otherwise, we have fixed $\alpha=0.5$ 
leading to pulse intensities $10^{3} \ldots
10^{4}$ ${\rm W}/{\rm cm}^2$. 
We apply a rapidly converging numerical 
scheme~\cite{zhu,janthesis} that has been implemented in
the {\tt octopus} code.~\cite{octopus}

\begin{figure}
\includegraphics[width=0.8\columnwidth]{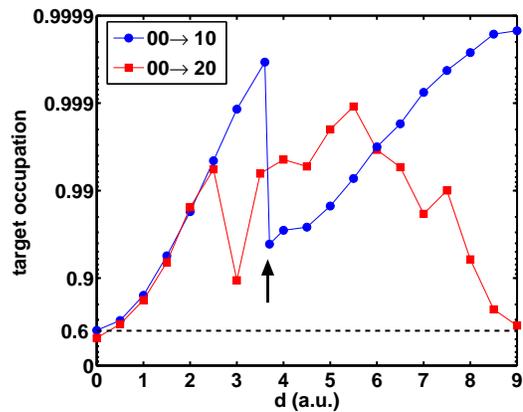}
\caption{(color online). Maximum occupation (logarithmic scale)
of the target state as a function of the interdot distance $d$ 
in transitions $|00\big>\rightarrow |10\big>$ with pulse length $T=50$ 
(circles), and $|00\big>\rightarrow |20\big>$ with $T=100$ 
(squares). The lines are to guide the eye.
The dashed line denotes the maximum
target-state occupation $0.6$ for
a single harmonic oscillator. The jump marked by an arrow
is due to a resonance effect (see text).
%The confinement strength of the double quantum dot
%is $\omega_0=0.5$.}
}
\label{max}
\end{figure}

We consider excitations only from the ground state and set
$|\Phi_{\rm I}\big>=|00\big>$. Figure~\ref{max}
shows the maximum overlaps $|\left<\Psi(T)|\Phi_{\rm F}\right>|^2$,
i.e., maximum occupations of the target 
states $|10\big>$ and $|20\big>$ as a 
function of the interdot distance (note the logarithmic scale). 
The pulse lengths are fixed to
$T=50$ and $T=100$, respectively.
When $d$ is increased, the target-state occupations increase 
from the HO value $\lesssim 0.6$~(see Ref.~[\onlinecite{qdcontrol}]) 
exponentially to close to one. As expected, at 
large interdot distances ($d\gtrsim 8$), corresponding to the uncoupling of 
the DQD, the occupations for $|00\big>\rightarrow |20\big>$ decrease 
back to the HO value marked by a dashed line. 
On the other hand, the occupation for 
$|00\big>\rightarrow |10\big>$ increases even further in
this limit due to the asymptotic degeneracy of the states (see above).
Generally, for this transition a pulse length of $T=50$ a.u. $\sim 3$ ps is 
sufficient to achieve high occupations. In the transition
$|00\big>\rightarrow |20\big>$ instead, $T=100$ a.u. $\sim 6$ ps
is needed for the same accuracy. 
%Nevertheless, the
%required frequencies are in the THz regime. 

As seen in Fig.~\ref{max}, the target-state 
occupations above $90\%$ are generally obtained 
at distances $2<d<8$, i.e., at $d\sim 20\ldots 80$ nm.
This length scale is well realizable in 
experiments.~\cite{pettascience,pettaPRL,gorman}
We point out that the most distinctive deviations 
in the occupations in this regime arise from
resonance effects. For example, the jump in
the $|00\big>\rightarrow |10\big>$ transition at
$d=3.6\ldots 3.7$ (see the arrow in Fig.~\ref{max}), 
where the occupation decreases from $0.9997$
to $0.9591$, is due to the degeneracy of $\omega_{00}^{10}$
and $\omega_{10}^{20}-\omega_{20}^{30}$. The degeneracy
disturbs the optimal transition path leading to reduced
maximum occupation. 
%Overall, however, very high occupations
%are obtained despite the 
%In this respect,
%the optimal transition path is sensitive to $d$

\begin{figure}
\includegraphics[width=0.85\columnwidth]{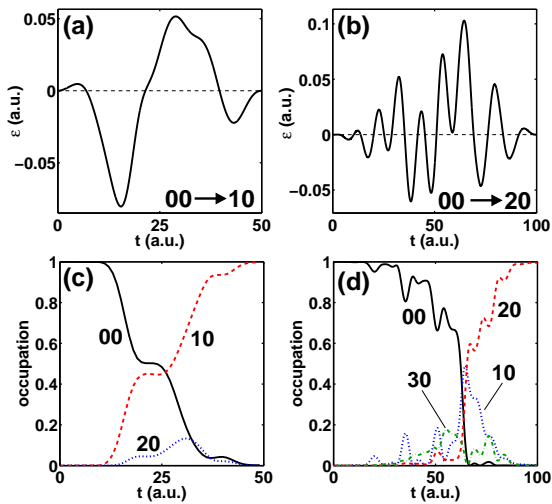}
\caption{(color online). Upper panel: Optimized pulses ({\em x} components) 
for transitions $|00\big>\rightarrow |10\big>$ (a)
and $|00\big>\rightarrow |20\big>$ (b). The interdot distances
are fixed to $d=3$ and $5$ and the pulse lengths 
to $T=50$ and $100$, respectively.
Lower panel: Occupations of states involved in
the transitions.
}
\label{pulses}
\end{figure}

In Fig.~\ref{pulses}
we give two examples of optimized pulses and the occupations
of states during transitions $|00\big>\rightarrow |10\big>$ 
when $d=3$ (a,c) and
 $|00\big>\rightarrow |20\big>$ when $d=5$ (b,d).
%The converged optimal pulses in (a) and (b) are linearly 
%polarized, so that in our coordinate system [see Eq.~(\ref{pot})]
%$\epsilon_x(t)=\epsilon_y(t)$.
%For clarity, we have excluded envelope functions
%(to start from zero amplitude) from the optimal pulses.
The pulse for $|00\big>\rightarrow |10\big>$ has the resonant
frequency $\omega_{00}^{10}$ as the major component,
but as seen in the occupations, state $|20\big>$ is also
populated during the transition. For comparison, we 
exposed the system also to a CW having the resonant (Rabi) 
frequency $\Omega_R=\omega_{00}^{10}$
and the same fixed length ($T=50$) as the optimized pulse.
We set the pulse amplitude to 
$\Omega_R/\mu_{00}^{10}=\pi/(\mu_{00}^{10}\,T)$ to
satisfy the $\pi$-pulse condition. This leads to
occupation $0.93$ of state $|10\big>$. 
In view of the $0.999$ occupation achieved by the 
optimal pulse, OCT is superior
to the CW approach.
A comprehensive comparison of occupations achieved
with different pulse lengths applying OCT and 
CWs for a quantum-ring system is given 
in Ref.~[\onlinecite{ring}].

As shown in Fig.~\ref{pulses}(d), the $|00\big>\rightarrow |20\big>$
transition must be mediated by the $|10\big>$ state
since a direct transition is forbidden by the 
dipole selection rules. However, the 
intermediate state $|10\big>$ does not need to get fully
populated during the optimized transition process. 
The higher states are also involved in the process, and
in this example the state $|30\big>$ receives
considerable occupation with a maximum of $\sim 20\%$ 
at $t\sim 90$. The final target-state occupation is 
$0.998$, whereas using two resonant CWs one after the 
other we could not exceed $0.9$ (pulse length fixed to $T=100$).
Generally,
in multilevel transitions of this type,
the efficiency of OCT is pronounced with respect
to CWs due to the multiplication of errors in the 
latter approach when
full population of intermediate states is required
before changing the resonant frequency.

The control of electron transport in the DQD
requires {\em initialization} of the state
by localizing the electron in the one of the dots.
Namely, the single-electron ground state of the DQD 
is a gerade state with 
half of the electron in one well and
half in the other (see the lowermost figure
in the right panel of Fig.~\ref{spectrum}).
Only in the limit $d\rightarrow \infty$ the
localized states become degenerate eigenstates.
%In nature, the symmetry between the dots is 
%inevitably broken due to, e.g., slight deviations 
%in the confining potential. 
However, the
initial state can be fixed with certainty 
by applying a constant external field or by
adjusting the gates in the DQD device in order
to create a potential shift between the two dots.
After the initial localization, electron transport
into another well can be driven 
using a linear switch or a CW with the resonant 
confinement frequency $\omega_0$. These types of
transport on time scales of $10^{-10}$ s, close
to the required times for SWAP operations in 
experiments,~\cite{pettascience}
were reported very recently by F{\o}rre and 
co-workers.~\cite{forre}

\begin{figure}
\includegraphics[width=0.9\columnwidth]{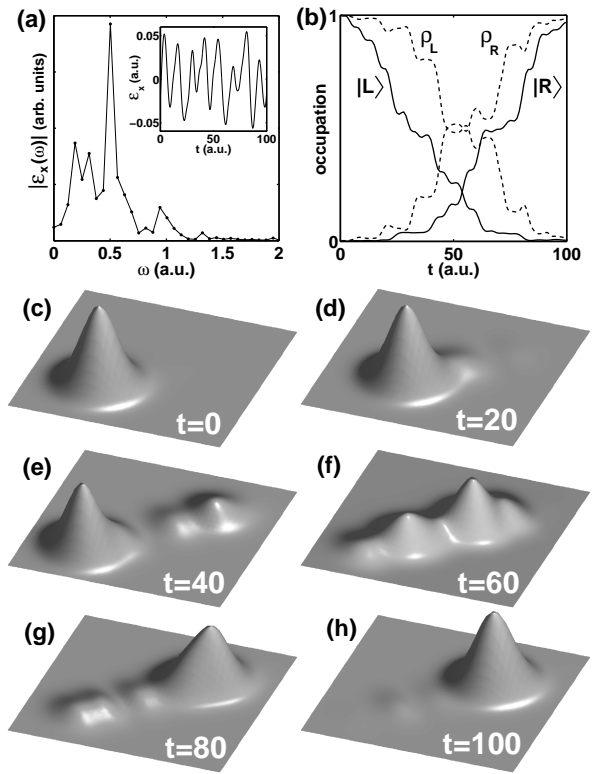}
\caption{(a) Spectrum of the optimized pulse (inset)
for the electron transport process $|L\big>\rightarrow |R\big>$
in a fixed time $T=100$. The pulse has a rectangular envelope
$A(t)=1$ and the penalty factor is $\alpha=1$.
(b) Occupations of states $|R\big>$ and $|L\big>$ (solid lines)
and the integrated densities $\rho_{\rm R}$ and $\rho_{\rm L}$ 
(dashed lines) in the right and
left dot during the transport.
(c-h) Snapshots of the total electron density 
$\rho_{\rm R}+\rho_{\rm  L}$. 
The double quantum dot has $d=6$,
$\omega_0=0.5$, and well-depth asymmetry of $V_0=0.2$.
}
\label{tunnel}
\end{figure}

Now we show that OCT provides a very fast (switching
times of a few picoseconds) 
and stable alternative to control the electron transport 
in a DQD. First we break the degeneracy of the ground state
by setting the external potential in the (lower) 
left dot $V_c^{\rm left}\rightarrow V_c^{\rm left}-0.2$.
Then the ground state $|00_{L}\big>=|L\big>$ and the first excited state 
$|00_{R}\big>=|R\big>$ correspond to electron localization in the (lower) 
left and (upper) right dot, respectively, provided
that $d$ is sufficiently large. The result of the
OCT calculation for transition $|L\big>\rightarrow |R\big>$
is shown in Fig.~\ref{tunnel}.
The pulse length is fixed to $T=100$ ($\sim 6$ ps), and the interdot distance
is $d=6$ corresponding to relatively large coupling between the dots.
The spectrum of the optimized pulse (a) has
a large peak at the resonant frequency 
$\omega=\omega_0=0.5$ a.u. $\sim 9$ THz,
and a few smaller peaks at $\omega\sim 0.2...0.3$ and
$\omega\sim 1$. The small peaks correspond to transitions
in the higher states which get significantly populated
in the transitions. This is visualized in Fig.~\ref{tunnel}(b)
showing the occupations of $|L\big>$ and $|R\big>$ (solid lines) that sum 
up only to about $50\%$ in the middle of the transition
at $T\sim 50$. We also plot the integrated electron densities
$\rho_{\rm R}$ and $\rho_{\rm L}$
(dashed lines) in the (lower) left and (upper) right parts of the DQD,
respectively.
The quasi-periodic oscillations in the densities indicate that
the electron charge transfers in blobs as visualized in the 
snapshots in Figs.~\ref{tunnel}(c-h). In this example
we find the final occupation 
$|\left<\Psi(T)|R\right>|^2=0.985$. It is worth noting that
the final occupations are closer to one at larger interdot distances 
so that the coupling between the dots is weaker. In this regime
the optimized pulses may attain a linear slope. This immediately
suggests that a combination of a linear field and OCT could 
be ideal in controlling transport in the uncoupled regime.
In this work, however, we focus on coupled DQDs which we in fact
find the most challenging in terms of electron control
in nanoscale applications.

There are three significant advantages in OCT with respect to a 
linear switch or a resonant CW when generating electron transport in
DQDs. First, the optimization procedure is insensitive to the interdot
coupling. The CW approach in principle requires an uncoupled system 
for electron transport with unit probability.~\cite{forre} We tested
a CW with the resonant $\omega_0$ frequency for the above example
with various amplitudes and could not exceed occupation $0.4$
of the target state $|R\big>$. Secondly, the optimized pulses can
be very short in order to achieve sufficient occupation, since
the higher-lying states are incorporated in the control problem.
In the above example the pulse length of $\sim 6$ ps
is more than by a factor of ten smaller than in linear and CW 
switches 
and well below the recently measured coherence times of hundreds 
of nanoseconds in DQDs.~\cite{gorman} Thirdly, and probably most 
importantly, the OCT approach is insensitive to deviations in
the external potential, since the specific shape of the 
DQD is taken into account explicitly through the external potential 
in the Hamiltonian. We tested this by adding a fourth-order 
anharmonic term in the external potential [Eq.~(\ref{pot})].
This results in a change in the optimal pulse shape, and
in an increase in the required pulse length (in the case of a positive
anharmonicity). However, we found no decrease in the obtained 
target-state occupation when using anharmonicities that lead to 
dramatic loss of accuracy in the CW approach.~\cite{forre}
In practical applications the exact shape of the external 
potential could be obtained by measuring the differential
conductance in a single-electron transport experiment, and
thereafter numerically solving the (inverse) Schr{\"o}dinger 
equation. This approach has been applied to modeling
external impurities inside single quantum dots.~\cite{impurity}

We expect that the experimental creation of optimal laser 
pulses such as presented in this work will be soon within reach 
of laser technologies. The intensities $10^{3} \ldots
10^{4}$ ${\rm W}/{\rm cm}^2$ required for DQD excitations
can be already obtained in the THz regime by high-power
free-electron lasers.~\cite{williams} 
On the other hand, shaping of picosecond THz pulses
has been recently made possible by employing transient polarization 
grating.~\cite{stepanov} Quantum cascade lasers~\cite{tonouchi} 
may also provide an applicable route for precise pulse shaping 
in the THz regime. The rapid developments in the THz 
laser technology~\cite{posthumus_reimann} will eventually 
lead to the combination of sufficient pulse power and accurate
manipulation of the pulse shape.

Finally we point out that in our future work we aim at combining
our approach with magnetic-field optimization which could allow
us to coherently control the spin state simultaneously with the 
electron localization.

%To conclude, we have shown that a two-dimensional double quantum 
%dot has completely controllable single-electron states at
%a large range of the interdot distances. Coherent control of the 
%excitations can be achieved using optimal-control theory that
%gives optimal laser fields for desired transitions. We obtain
%very high occupations of the target states with laser
%pulses of only a few picoseconds. Controlled electron
%transport is achieved after initial localization, 
%and the optimized transport process is rather insensitive to the
%interdot coupling and to anharmonicities in the external
%potential. This opens up possibilities in coherent
%charge manipulation by terahertz laser pulses that
%could be refined using quantum-cascade techniques.

\begin{acknowledgments}
This work was supported by the EU's Sixth Framework
Programme through the Nanoquanta NoE 
(No. NMP4-CT-2004-500198), SANES project
(No. NMP4-CT-2006-017310), DNA-NANODEVICES
(No. IST-2006-029192), Barcelona Supercomputing
Center, the Humboldt Foundation, the Academy of
Finland, 
the Finnish Academy of Science and Letters
through the Viljo,
Yrj{\"o} and Kalle V{\"a}is{\"a}l{\"a} Foundation,
and the Deutsche Forschungsgemeinschaft through SFB
658.
\end{acknowledgments}


\begin{thebibliography}{ll}

\bibitem{pettascience} J. R. Petta, A. C. Johnson, J. M. Taylor,
E. A. Laird, A. Yacoby, M. D. Lukin, C. M. Marcus, M. P. Hanson, and
A. C. Gossard, 
Science, {\bf 309} 2180 (2005).

\bibitem{pettaPRL} J. R. Petta, A. C. Johnson, 
C. M. Marcus, M. P. Hanson, and A. C. Gossard,
Phys. Rev. Lett. {\bf 93}, 186802 (2004). 

\bibitem{gorman}
J. Gorman, D. G. Hasko, and D. A. Williams,
Phys. Rev. Lett. {\bf 95}, 090502 (2005). 

\bibitem{forre}
M. F{\o}rre, J. P. Hansen, V. Popsueva, and A. Dubois,
Phys. Rev. B {\bf 74}, 165304 (2006). 

\bibitem{garcia}
I. Grigorenko, O. Speer, and M. E. Garcia,
Phys. Rev. B {\bf 65}, 235309 (2002).

\bibitem{kosionis}
S. G. Kosionis, A. F. Terzis, and E. Paspalakis,
Phys. Rev. B {\bf 75}, 193305 (2007).

\bibitem{harju}
A. Harju, S. Siljam{\"a}ki, and R. M. Nieminen,
Phys. Rev. Lett. {\bf 88}, 226804 (2002).

\bibitem{wensauer}
A. Wensauer, O. Steffens, M. Suhrke, and U. R{\"o}ssler, 
Phys. Rev. B {\bf 62}, 2605 (2000).

\bibitem{oct} A. P. Peirce, M. A. Dahleh, and
H. Rabitz, Phys. Rev. A {\bf 37}, 4950 (1988);
R. Kosloff, S. A. Rice, P. Gaspard, S. Tersigni, and D. J. Tannor,
Chem. Phys. {\bf 139}, 201 (1989).

%\bibitem{computer} See, e.g.,
%D. P. Di Vincenzo and C. H. Bennett, Nature (London)
%{\bf 404}, 247 (2000).

\bibitem{qdcontrol} See, e.g., A. G. Butkovskiy and Y. I. Samoilenko
{\em Control of Quantum-Mechanical Processes and Systems},
(Kluwer, Dordrecht, 1990); 

\bibitem{schirmer} S. G. Schirmer, H. Fu, and A. I. Solomon,
Phys. Rev. A {\bf 63}, 063410 (2001).

\bibitem{ring} E. R{\"a}s{\"a}nen,
A. Castro, J. Werschnik, A. Rubio, and E. K. U. Gross, 
Phys. Rev. Lett. {\bf 98}, 157404 (2007).

\bibitem{zhu} W. Zhu and H. Rabitz, J. Chem. Phys. {\bf 109}, 385 (1998).

%\bibitem{sundermann} K. Sundermann and R. de Vivie-Riedle, 
%J. Chem. Phys. {\bf 110}, 1896 (1999).

\bibitem{janthesis} J. Werschnik, {\em Quantum Optimal Control Theory: 
Filter Techniques, Time-Dependent Targets, and Time-Dependent 
Density-Functional Theory}, (Cuvillier, G{\"o}ttingen, 2006).

\bibitem{octopus} A. Castro,
H. Appel, M. Oliveira, C. A. Rozzi, X. Andrade,
F. Lorenzen, M. A. L. Marques, E. K. U. Gross, and A. Rubio,
Phys. Stat. Sol. (b) {\bf 243}, 2465 (2006).

\bibitem{impurity}
E. R{\"a}s{\"a}nen, J. K{\"o}nemann, R.~J. Haug,
M.~J. Puska, and R.~M. Nieminen, Phys. Rev. B {\bf 70},
115308 (2004).

\bibitem{williams} G. W. Williams, Rep. Prog. Phys. 
{\bf 69} 301 (2006).

\bibitem{stepanov} A. G. Stepanov, J. Hebling, and
J. Kuhl, Opt. Express {\bf 12}, 4650 (2004).

\bibitem{tonouchi} M. Tonouchi, Nature Photonics {\bf 1}, 97 (2007).

\bibitem{posthumus_reimann} For reviews, see, e.g., 
J. H. Posthumus, Rep. Prog. Phys. {\bf 67}, 623 (2004);
K. Reimann, Rep. Prog. Phys. {\bf 70} 1597 (2007).

\end{thebibliography}
\end{document}